# Anisotropic Pauli spin-blockade effect and spin-orbit interaction field in an InAs nanowire double quantum dot

Ji-Yin Wang,[†] Guang-Yao Huang,[†] Shaoyun Huang,[*,†] Jianhong Xue,[†] Dong Pan,[‡] Jianhua Zhao,[*,‡] and H. Q. Xu[*,†,§]

[†] Beijing Key Laboratory of Quantum Devices, Key Laboratory for the Physics and Chemistry of Nanodevices, and Department of Electronics, Peking University, Beijing 100871, China

[‡] State Key Laboratory of Superlattices and Microstructures, Institute of Semiconductors, Chinese Academy of Sciences, Beijing 100083, China

[§] Division of Solid State Physics, Lund University, Box 118, S-22100 Lund, Sweden

## Abstract

We report on experimental detection of the spin-orbit interaction field in an InAs nanowire double quantum dot device. In the spin blockade regime, leakage current through the double quantum dot is measured and is used to extract the effects of spin-orbit interaction and hyperfine interaction on spin state mixing. At finite magnetic fields, the leakage current arising from the hyperfine interaction is suppressed and the spin-orbit interaction dominates spin state mixing. We observe dependence of the leakage current on the applied magnetic field direction and determine the direction of the spin-orbit interaction field. We show that the spin-orbit field lies in a direction perpendicular to the nanowire axis but with a pronounced off-substrate-plane angle. It is for the first time that such an off-substrate-plane spin-orbit field in an InAs nanowire has been detected. The results are expected to have an important implication in employing InAs nanowires to construct spin-orbit qubits and topological quantum devices.

*Correspondence should be addressed to Professor H. Q. Xu (hqxu@pku.edu.cn), Dr. Shaoyun Huang (syhuang@pku.edu.cn), or Professor Jianhua Zhao (jhzhao@semi.ac.cn).

Spin-orbit interaction couples the spin magnetic moment of an electron to its orbital degree of freedom. Because of its potential applications in the development of novel all-electrical controlled spintronic devices and spin-orbit qubits[1-3] and its fundamental role in the formation of new topological states of matter such as topological insulators and topological superconductors,[4, 5] spin-orbit interaction in solid state systems has attracted great interests in recent years. Semiconductor InAs and InSb nanowires have been demonstrated to exhibit strong spin-orbit interaction.[6-8] Based on these nanowires, advanced quantum devices have been developed, including quantum dot (QD) devices, in which fast electrical manipulations of electron spins have been achieved[9,10] and semiconductor-superconductor hybrid devices, in which Majorana zero modes could be created and manipulated.[11-15] Of great importance for future development of quantum computation technology based on semiconductor nanowires QDs and topological superconducting nanowires is to know in great detail the orientation of the spin-orbit interaction field and find efficient ways of manipulating spins and achieving quantum phase transitions in semiconductor nanostructures, including most promising systems of InAs and InSb nanowires. The anisotropic effects of spin-orbit interaction in GaAs/AlGaAs heterostructure QDs have recently been investigated,[16, 17] where the direction of the spin-orbit interaction field with respect to the crystallographic axes was analyzed. Anisotropic effects of spin-orbit interaction have also been observed in double quantum dots (DQDs) made from InAs and InSb nanowires.[18, 19] However, an experimental determination of the orientation of the spin-orbit interaction field in these nanowire quantum structures still remains challenging.

In this Letter, we report an experimental detection of the spin-orbit interaction field and its interplay with hyperfine interaction in an InAs nanowire DQD. In the spin blockade regime, the leakage current through the DQD is measured and the roles of spin-orbit interaction and hyperfine interaction on spin state mixing are extracted. We find that the hyperfine interaction induces spin state mixing mainly at small, around zero magnetic fields, while the spin-orbit interaction is dominant at finite magnetic fields. By applying the magnetic field along different directions with respect to the nanowire axis, the leakage current is found to depend on the direction of the applied magnetic field. Based on these measurements, the direction of the spin-orbit interaction field is extracted. We show that the spin-orbit field is perpendicular to the nanowire, consistent with previous reports, but possesses a pronounced off-substrate-plane angle. Determination of such an off-substrate-plane oriented spin-orbit field in a semiconductor nanowire could be very important for constructing spintronic and spin-orbit qubit devices and for realizing a topological superconducting nanowire in which Majorana zero modes can exist and be employed for topological quantum computation.

The device is fabricated from a single-crystalline pure-phase InAs nanowire with a diameter of ~35 nm on a Si/SiO$_2$ substrate. The nanowires are grown on a Si (111) substrate via molecular-beam epitaxy (MBE) using Ag particles as seeds.[20] Figure 1a shows a scanning electron microscope (SEM) image of a fabricated InAs nanowire device. The device fabrication starts from transferring MBE grown InAs nanowires from the growth substrate onto the Si/SiO$_2$ substrate with predefined markers. After a nanowire is selected and located with respect to the markers on the Si/SiO$_2$ substrate, source and drain contact electrodes are fabricated by pattern definition via electron-beam lithography (EBL), deposition of a layer of 5 nm thick titanium and 90 nm thick gold via electron-beam evaporation (EBE), and lift-off process. Here, it should be noted that right before the metal deposition, the contact regions are etched in diluted (NH$_4$)$_2$S$_x$ solution to remove the oxide layer formed on the nanowire surface.[21] Subsequently, the InAs nanowire is covered by a layer of 10 nm thick HfO$_2$ via atomic layer deposition (ALD). An array of local finger gates with a width of 30 nm and an array pitch of 100 nm is fabricated on top of the HfO$_2$ covered nanowire by using EBL, EBE and lift-off process again. Transport measurements of the fabricated device are performed in a dilution refrigerator equipped with a vector magnet at a temperature of 15 mK.

Figure 1b shows a schematically cross-sectional view of the device and the measurement circuit setup in which a source-drain bias voltage $V_{SD}$ is applied to the source with the drain grounded. The silicon substrate works as a global back gate to which a voltage $V_{BG}$ is applied to tune the electrostatic potential to the Fermi level in the whole nanowire. We note that throughout this work, $V_{BG}$ has been fixed at 6 V to set the Fermi level within the conduction band of the nanowire before top gates are used. Local finger gates G3 to G6 have been used to build a DQD in the nanowire. The gates G3 and G6 are assigned to generate two outer barriers. The gates G4 and G5 are used to tune the electrostatic potentials of each QD. Simultaneously, the gates G4 and G5 also work to control inter-dot coupling. Such tuning scheme ensures that the electrons residing in the two QDs can be emptied before the barriers become too opaque to detect the current flow.[22] In order to avoid undesired barriers formed in other segments of the nanowire, positive voltages 0.5, 0.5 and 0.2 V are applied to gates G1, G2 and G7, respectively.

Figure 1c shows the source-drain current $I_{SD}$ measured as a function of local gate voltages $V_{G4}$ and $V_{G5}$ (charge stability diagram) for the DQD defined by setting local gate voltages $V_{G3} = -0.33$ V and $V_{G6} = -0.30$ V at $V_{SD} = 2$ mV. The notation ($N_L$, $N_R$) in the figure indicates electron numbers in the left and right QDs. The empty (0,0) charge state can be certainly assigned because no more lower charge state is observed at bias voltage as high as $V_{SD} = 30$ mV, which is much larger than the addition energies of two QDs (see details in Supporting Information). The charging energy $E_C^{L(R)}$ of the left (right) QD in the few electron regime is extracted to be ~5.7 (~4.4) meV and the first orbital quantization energy $E_{orb}^{L(R)}$ of the left (right) QD is extracted to be ~6.5 (~2.1) meV (see details in Supporting Information), showing that the DQD is asymmetric and the two QDs are confined differently. In Figure 1c, the yellow rectangle marks the region of charge transitions in the two electron regime, where current can flow through the DQD via a cycle of electron transfers (1,0)→(1,1)→(2,0)→(1,0) or hole transfers (2,1)→(1,1)→(2,0)→(2,1).[23] When state (1,1) is a triplet and triplet state (2,0) is not accessible in energy, the charge transition (1,1)→(2,0) is forbidden in both the electron and the hole transfer processes owing to the Pauli spin blockade.[24] The blockade can be lifted by spin mixing due to hyperfine interaction and spin-orbit interaction, hence giving rise to finite leakage current in the blockade region.[6, 22, 25] In this letter, we focus on the leakage current in this region to extract the intrinsic properties of the hyperfine interaction and spin-orbit interaction.

Figure 2a shows the energy diagram of the (1,1) and (2,0) states in a magnetic field $\vec{B}$ and spin-nonconserving coupling induced by hyperfine interaction. The $T_+(1,1)$ and $T_-(1,1)$ states are separated away from the S(1,1) and $T_0(1,1)$ states by Zeeman energy $E_Z = g^*\mu_B|\vec{B}|$. The gray stripe denotes the characteristic energy scale $E_N$, over which hyperfine interaction takes effects. The $E_N$ can be determined as $g^*\mu_B B_N$, where $B_N$ is the root mean square of randomly distributed nuclear fields. At small external magnetic field ($B < B_N$), all the three (1,1) triplet states can be mixed with the S(1,1) state by the dynamic fluctuations of nuclear fields in the two QDs. When external magnetic field is large enough ($B > B_N$), the $T_\pm(1,1)$ states are separated away from the gray-stripe zone and the mixing of the $T_\pm(1,1)$ states with the S(1,1) state via the nuclear fields is suppressed. As a result, the hyperfine-interaction induced leakage current decreases dramatically with increasing external magnetic field.[26] Figure 2b shows the energy diagram of the relevant spin states and spin-nonconserving transitions induced by spin-orbit interaction in a finite magnetic field. Here, again, the $T_\pm(1,1)$ states are separated away because of Zeeman splitting. However, due to the presence of spin-orbit interaction, the S(1,1) and $T_0(1,1)$ states are coupled to form one spin blockaded state, which we label as state α, and one non-blockaded state, which is labeled as state β. Thus, around zero magnetic field, only the β state can decay to the S(2,0) state and the other three states are forbidden to transit into the S(2,0) state. When an external magnetic field is applied, three (1,1) states, namely the $T_+$, $T_-$ and β states are able to transit to the S(2,0) state as shown in Figure 2b. As a result, the leakage current is enhanced correspondingly with increasing magnetic field.[27] The spin-orbit induced leakage current strongly depends on the direction of the external magnetic field.[28] In a simple physical picture, the spin-orbit field $\vec{B}_{SO}$ causes the spin to precess around the external magnetic field $\vec{B}$. The spins are rotated and spin-flip tunneling takes place to lift the spin blockade. The precession angle and the induced leakage current are maximum when $\vec{B}_{SO}$ is perpendicular to $\vec{B}$, while the parallel part of $\vec{B}_{SO}$ does not contribute to spin mixing. Thus, the spin-orbit interaction induced leakage current is anisotropic and is dependent on the direction of $\vec{B}_{SO}$ with respect to $\vec{B}$.

Figure 2c shows the enlarged region marked by the yellow rectangle in Figure 1c. Since the bias window is smaller than the first orbital quantization energy of each QD, only the first orbital of each QD is accessible in energy and spin blockade takes place in the whole triangle regions marked by the dashed lines in Figure 2c. On the contrary, finite leakage current through the DQD is clearly observed in the triangle regions owing to spin mixing mainly induced by hyperfine interaction and spin-orbit interaction.[26, 27] Figure 2d shows the same triangle region of Figure 2c with applying an external magnetic field of 20 mT perpendicular to the substrate. Compared with that in Figure 2c, the leakage current seen in Figure 2d is significantly enhanced around the baselines (indicated by blue arrows) of the triangles. As discussed in Figures 2a and 2b, the reason for the change in leakage current in a magnetic field is that the roles played by hyperfine interaction and spin-orbit interaction on spin mixing and spin flips are influenced by the external magnetic field.

Figure 3 shows the leakage current through the DQD in the spin blockade regime as a function of detuning energy and external magnetic field applied along different directions. Here, the detuning energy $\varepsilon$ is tuned by sweeping $V_{G4}$ and $V_{G5}$ simultaneously along the yellow arrow line in Figure 2c. In Figure 3, the external magnetic field is applied along the default coordinate axes of the magnets, $x'$, $y'$ and $z'$. There is a misalignment of ~15° between the magnet axis $x'$ and the nanowire axis. Figure 3a shows the measured leakage current $I_{SD}$ as a function of detuning energy and magnetic field applied along the $y'$ axis. Persistent leakage current is observed around zero detuning energy, while the leakage current is merely visible around zero magnetic field at finite detuning energy. A linecut along $\varepsilon = 0$ marked by the black bar is plotted as opened black squares in Figure 3d. It can be seen that the current profile is composed of a narrow peak around $|\vec{B}| = 0$ and a broad dip at relatively larger magnetic fields, reflecting the presence of two different spin nonconserving transport mechanisms. According to Figures 2a and 2b, we infer that hyperfine interaction prevails in spin mixing around zero magnetic field but its effect reduces dramatically at finite magnetic fields, giving rise to a peak in the leakage current lineshape. This agrees well with the results observed in GaAs/AlGaAs[29] and InGaAs/InP heterostructure DQDs.[30] On the other hand, the broad dip in the leakage current lineshape in Figure 3d arises from spin state mixing induced by spin-orbit interaction. Similar behaviors have been observed in GaAs/AlGaAs heterostructure[31] and silicon DQDs.[32]

Theoretically, the leakage current $I_{\text{Leak}}$ through a DQD in the Pauli spin blockade regime at a finite magnetic field $|\vec{B}|$ can be estimated from[26, 27]

$$I_{\text{Leak}} = I_{\text{HF}}^0 S\left(\frac{\sqrt{3}|\vec{B}|}{B_{\text{N}}}\right) + I_{\text{SO}}^0 \left(1 - \frac{8B_{\text{C}}^2}{9\left(|\vec{B}|^2 + B_{\text{C}}^2\right)}\right) + I_{\text{B}}. \qquad (1)$$

The first term on the right side represents the hyperfine-interaction induced leakage current, where $S(x)$ is a special function defined in Ref. 26. The second term on the right side describes the leakage current resulting from spin-orbit interaction, which depicts a Lorentzian shaped dip with a width $B_{\text{C}}$, where $B_{\text{C}}$ is related to the strength of spin-orbit interaction (see Ref. 27). The last term $I_{\text{B}}$ on the right side is the overall background current caused by the other spin mixing mechanisms, such as spin-flip co-tunneling.[33]

The red-solid line shown in Figure 3d is the result of fitting the measured data to Eq. (1). Clearly, the theory agrees well with experiment. From the fit, we can extract an effective nuclear field of ~5.6 mT, which is consistent with previous works.[6, 22] The value of $B_{\text{C}}$ is ~16.5 mT and the overall background current $I_{\text{B}}$ is ~0.3 pA. Figures 3b and 3c represent the leakage current measured as a function of the detuning energy and the magnetic field applied along the *x′* and *z′* axis, respectively. Figures 3e and 3f show the plots of the leakage current as a function of the magnetic field applied along the two directions at zero detuning energy (i.e., along the line cuts of $\varepsilon = 0$ marked by black bars). In both directions, the leakage current exhibits a single dip lineshape, implying that spin-orbit interaction dominates the spin mixing.[19] These data can also be well fitted using Eq. (1) and the extracted value of $B_{\text{C}}$ is ~9.0 mT in Figure 3e and ~8.5 mT in Figure 3f (see details in Supporting Information). Overall, the values of $B_{\text{C}}$ are different in the measurements with the magnetic field applied along the three directions, indicating that the degree of spin state mixing by spin-orbit interaction is anisotropic.[31] In particular, among the three extracted values, $B_C$ is the largest when the magnetic field is applied along the $y'$ axis, implying that the effective spin-orbit interaction field is oriented closely along the $y'$ direction. Figures 3g, 3h and 3i show the leakage current measured at a fixed detuning energy $\varepsilon = 235$ μeV, i.e., along the line cuts marked by green bars in Figures 3a, 3b and 3c, as a function of magnetic field applied along the $y'$, $x'$ and $z'$ directions. Compared with the results measured at zero detuning energy shown in Figures 3d, 3e and 3f, the leakage current seen in Figures 3g, 3h and 3i becomes smaller. This is because of strong reductions in coupling between the (1,1) states and the S(0,2) state at the finite detuning energy. In Figure 3g, the leakage current exhibits a single-peak lineshape, indicating that in this case the leakage current induced by the hyperfine interaction overwhelms that induced by the spin-orbit interaction. In Figures 3h and 3i, a dip-like lineshape of the leakage current is still observed at the finite detuning energy, implying that the leakage current induced by the spin-orbit interaction remains dominant when magnetic field is applied along the $x'$ and $z'$ directions.

Figure 4 shows the leakage current through the DQD measured at detuning $\varepsilon = 70$ μeV with the magnetic field rotated in three orthogonal planes. Here, the magnet coordinate system is rotated by 15° around the $z'$ axis so that in the new coordinate system the nanowire is aligned with the $x$ axis. The measurements are performed with a magnetic field of $|\vec{B}| = 20$ mT, which is significantly larger than $B_N$, being rotated in the $xy$, $xz$ and $yz$ planes in order to determine the direction of the effective spin-orbit interaction field $\vec{B}_{SO}$. It is seen that the leakage current exhibits a strong angular dependence when the magnetic field is applied in both the $xy$ and $yz$ planes, but it is not when the magnetic field is rotated in the $xz$ plane. As seen in Figure 4a, the leakage current has a minimum when the magnetic field is oriented nearly perpendicular to the nanowire axis in the $xy$ plane. Thus, the in-plane component of the effective spin-orbit field is aligned perpendicular to the nanowire axis. Similarly, in Figure 4c a minimum leakage current is found when the magnetic field is applied in the *yz* plane, showing that the effective spin-orbit field has a large perpendicular component with respect to the nanowire axis. However, it is seen that this perpendicular component does not point exactly along the *y* axis. In Figure 4b, the leakage current shows a weak dependence on the magnetic field direction, implying that the projection component of the spin-orbit field in the *xz* plane is negligibly small, if it is not zero.

The measured leakage current $I_{\text{Leak}}$ at the finite detuning energy and the finite magnetic field can be expressed as

$$I_{\text{Leak}} = I_{\text{SO}}^0 \left[ 1 - \frac{8}{9} \frac{(\frac{B_C^0}{\sin \alpha})^2}{|\vec{B}|^2 + (\frac{B_C^0}{\sin \alpha})^2} \right] + I_B', \qquad (2)$$

where $B_C^0$ is a constant and $\alpha$ is the angle between $\vec{B}$ and $\vec{B}_{\text{SO}}$. The first term on the right side is the leakage current contributed by the spin-orbit interaction. The second term on the right side $I_B'$ is the leakage current caused by the spin state mixing mechanisms other than the spin-orbit interaction and is assumed independent of the magnetic field direction. Now, we fit the measurements of the leakage current with the magnetic field applied in the *xy* and *yz* planes to Eq. (2). The red-solid lines in Figures 4a and 4c are the results of the fits, which show excellent agreement with the experimental results.

A detailed analysis of the results presented in Figure 4 allows us to extract the direction of the effective spin-orbit field $\vec{B}_{\text{SO}}$ experienced by electrons that pass through the middle tunneling barrier in the DQD device. The effective spin-orbit field $\vec{B}_{\text{SO}}$ is found to be in a direction pointing nearly perpendicular to the nanowire axis but not in the substrate plane (i.e., the *xy* plane). In detail, the extracted spin-orbit field $\vec{B}_{\text{SO}}$ has an azimuth angle of ϕ~88° from the *x* axis and a polar angle of θ~75° from the *z* axis. The ~2° deviation from the 90° perpendicular direction is most likely due to the uncertainty presented in the measurements and analysis. However, the extracted spin-orbit field $\vec{B}_{\text{SO}}$ has a noticeable angle of ~15° off from the substrate plane. We have also measured the magnetic-field-direction dependence of the leakage current at a larger detuning energy of $\varepsilon = 235$ μeV and found a similar off-plane angle of the spin-orbit field $\vec{B}_{SO}$ (see details in Supporting Information). It is for the first time that such an off-plane direction of the spin-orbit field $\vec{B}_{SO}$ in an InAs nanowire quantum device has been detected. The off-plane angle of the spin-orbit field $\vec{B}_{SO}$ could most likely arise due to the presence of the Dresselhaus spin-orbit interaction in the InAs nanowire employed in the fabrication of the DQD device. This is because the nanowire chip we have used for this work was shown to contain a substantial portion of pure-phase zincblende InAs nanowires grown along a <211> or <221> crystallographic direction[20,34,35] and for a zincblende InAs nanowire grown along a <211> or <221> crystallographic direction, a Dresselhaus spin-orbit field lying in the nanowire cross-sectional plane but pointing to an off-substrate-plane direction can be presented. Another source for the presence of the off-plane orientation of the spin-orbit field $\vec{B}_{SO}$ could arise from a potential asymmetry in the planar direction perpendicular to the nanowire axis, derived from the local gate structure of the device as seen in Figure 1a.

In conclusion, the leakage current through an InAs nanowire DQD device in the spin blockade regime has been measured and the effects of fluctuated nuclear fields and spin-orbit interaction field in the device have been determined. The root mean square of the nuclear fields is extracted to be ~5.6 mT, leading to an appearance of a noticeable leakage current only at small, around zero applied magnetic fields. The leakage current induced by the spin-orbit interaction is found to be dominant at finite magnetic fields and is applied-magnetic-field direction dependent. Based on the anisotropic properties of the leakage current, the direction of the spin-orbit interaction field has been extracted. The spin-orbit field is found to point to a direction perpendicular to the nanowire axis, but with a pronounced off-substrate-plane angle. It is for the first time that such an off-substrate-plane spin-orbit field in an InAs nanowire has been detected. The results may have important implications in constructing and manipulating spin-orbit qubits and Majorana zero modes using semiconductor nanowires.

**ACKNOWLEDGEMENTS**

This work is supported by the Ministry of Science and Technology of China (MOST) through the National Key Research and Development Program of China (Grant Nos. 2016YFA0300601, 2017YFA0303304, 2016YFA0300802 and 2017YFA0204901) and the National Natural Science Foundation of China (Grant Nos. 91221202, 91421303, 11274021, and 61504133). HQX also acknowledges financial support from the Swedish Research Council (VR).

**Figure Captions:**

**Fig. 1.** (a) Scanning electron microscope image of an InAs nanowire device made on the same chip with the same device structure as the one measured for this work. The nanowire has a diameter of ~35 nm, and the source and drain contact electrodes made on to the nanowire are a double metal layer of 5-nm-thick Ti and 90-nm-thick Au. The array of local finger gates is fabricated on top of the nanowire have a gate width of 30 nm and a pitch of 100 nm. A layer of 10-nm-thick $HfO_2$ is used as the gate dielectric. (b) Cross-sectional schematic view of the device and measurement circuit setup. Gates G3 and G6 are used to define the DQD, while gates G4 and G5 are used to control electron occupation and inter-dot coupling. A bias voltage $V_{SD}$ is applied to the source electrode and the drain electrode is grounded. A voltage $V_{BG}$ applied to the Si substrate (the back gate) is used to control the chemical potential in the nanowire. (c) Source-drain current $I_{SD}$ measured for the DQD as a function of $V_{G4}$ and $V_{G5}$ at $V_{SD} = 2$ mV. Other gate voltages used to defined the DQD are $V_{BG}$=6 V, $V_{G3}$=-0.33 V and $V_{G6}$=-0.30V. The notation ($N_L$, $N_R$) denotes electron numbers in the left and right QDs. The region marked by the yellow rectangle is investigated in detail in the following.

**Fig. 2** (a) Schematics for two-electron spin states and spin-nonconserving transitions caused by hyperfine interaction in the DQD in a finite magnetic field. The gray stripe indicates the characteristic energy scale $E_N$, over which spin state mixing by hyperfine interaction is effective. $E_Z$ is the Zeeman energy and $\varepsilon$ is the detuning energy. (b) Schematics for two-electron spin states in the presence of spin-orbit interaction and spin-nonconserving transitions in the DQD in a finite magnetic field. T-, β and T+ represent three (1,1) states that can decay to S(2,0) states and α denotes the remaining (1,1) state that is spin blockaded. $E_Z$ and $\varepsilon$ have the same meanings as in (a). (c) Zoom-in plot of the measurements in the region marked by the yellow rectangle in Figure 1c. (d) Corresponding measurements made in the same region as in (c) but with a magnetic field of 20 mT applied perpendicular to the substrate.

**Fig. 3** (a-c) Source-drain current $I_{SD}$ measured for the DQD as a function of detuning energy $\varepsilon$ and applied magnetic field $B$. (a), (b), and (c) are for the measurements (color plots) with $B$ applied along the magnet $y'$, $x'$, and $z'$ axis, respectively, as shown in the schematics depicted in the top panels. The nanowire is in the $x'y'$ plane (substrate plane) with a misalignment of ~15° from the $x'$ axis. The detuning energy $\varepsilon$ is tuned by sweeping $V_{G4}$ and $V_{G5}$ simultaneously along the yellow arrow in Figure 2c. (d) Source-drain current $I_{SD}$ measured as a function of magnetic field at zero detuning energy, i.e., linecut of the measurements along zero detuning energy $\varepsilon$ indicated by the black bars in (a) The red-solid line represents the best fit to the theory Eq. (1). (e) The same as in (d) but for $B$ applied along the $x'$ axis. (f) The same as in (d) but for $B$ applied along the $z'$ axis. (g-i) The same as in (d-f) but for the linecuts of measurements made at detuning energy $\varepsilon = 235$ μeV indicated by the green bars in (a-c). We note that the data in (d-i) have been symmetrized with respect to zero magnetic field.

**Fig. 4** (a) Magnetic-field direction dependent measurements of source-drain current $I_{SD}$ with the applied magnetic field $\vec{B}$ rotated in the $xy$ plane as depicted in the top panel. ϕ is the angle between $\vec{B}$ and the $x$ axis in the $xy$ plane. (b) The same as in (a) but for $\vec{B}$ rotated in the $xz$ plane. δ is the angle between $\vec{B}$ and the $z$ axis in the $xz$ plane. (c) The same as in (a) but for $\vec{B}$ rotated in the $yz$ plane. θ is the angle between $\vec{B}$ and the $z$ axis in the $yz$ plane. In all the measurements, the nanowire is aligned with the $x$ axis, the strength of the magnetic field is fixed at 20 mT, and the detuning energy is fixed at 70 μeV. The red-solid lines in (a) and (c) depict the best fits to the theory Eq. (2).

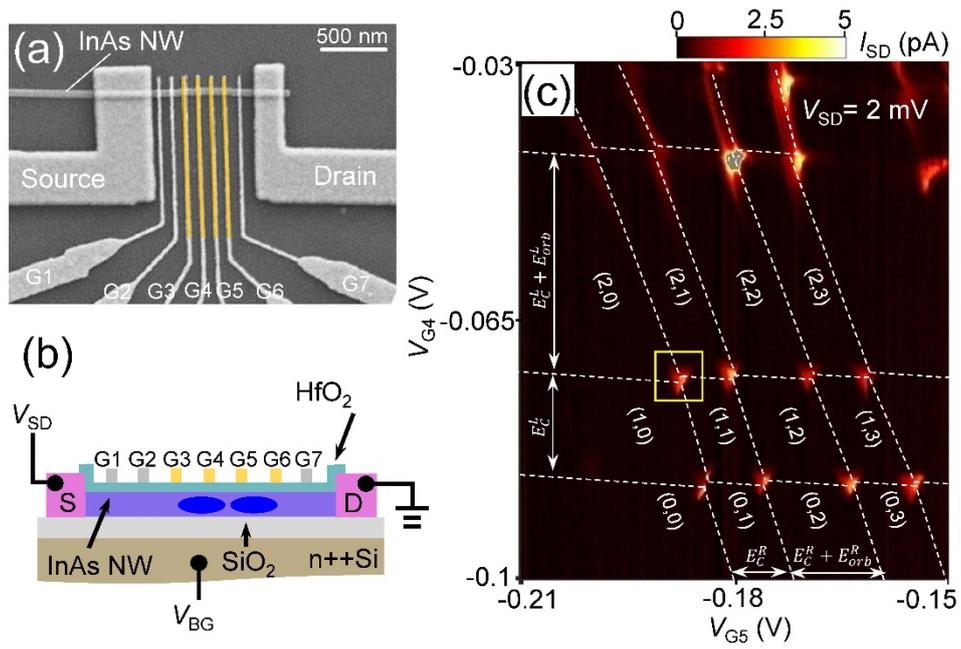

Fig. 1 by Ji-Yin Wang et. al

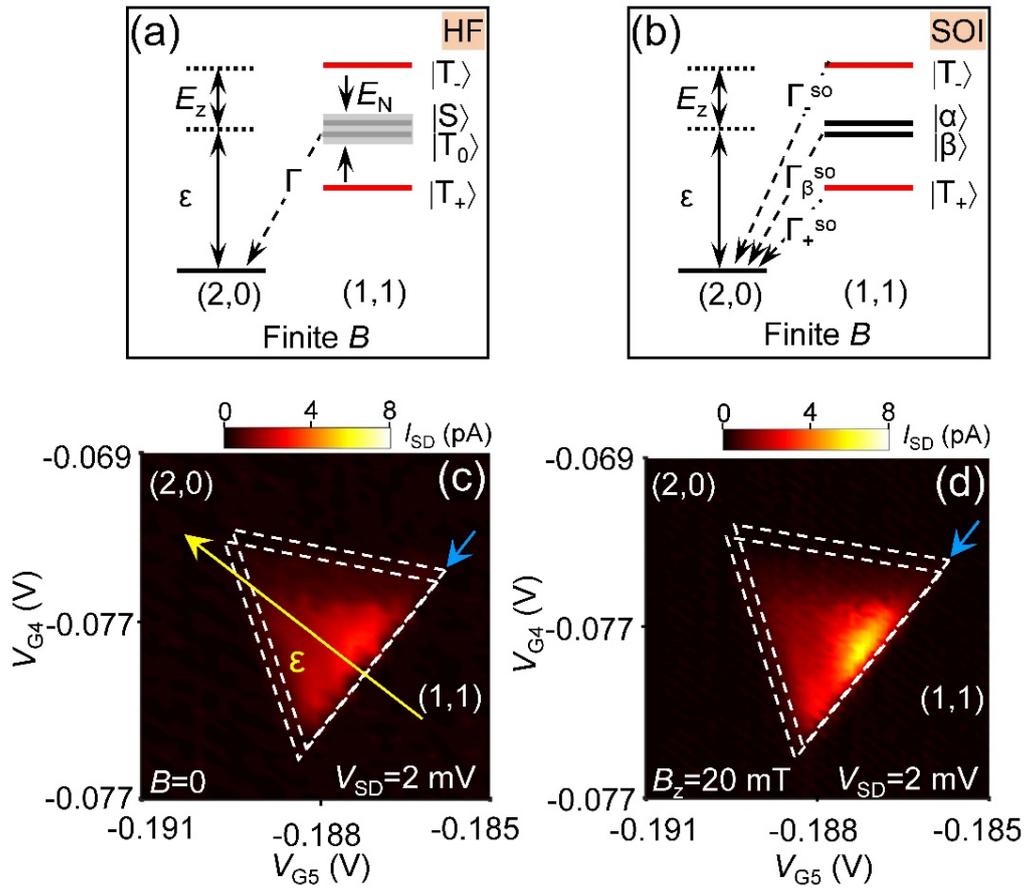

Fig. 2 by Ji-Yin Wang et. al

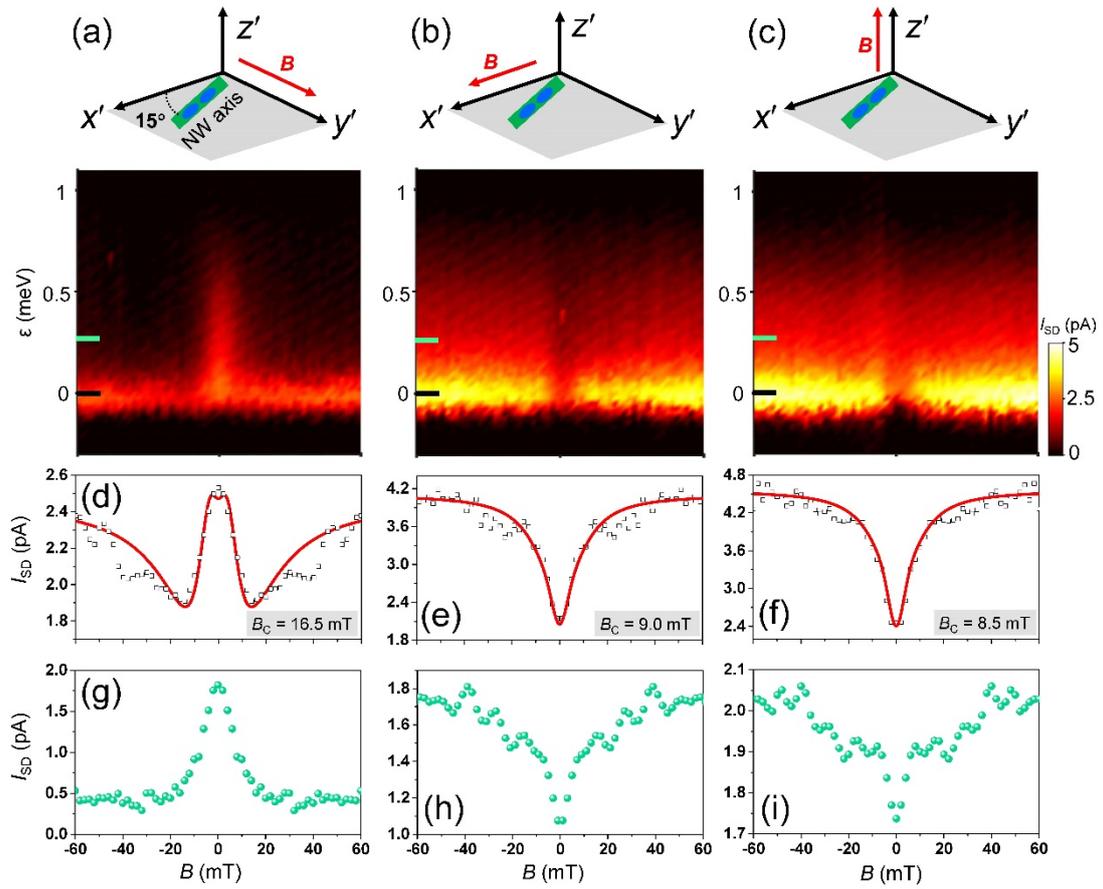

Fig. 3 by Ji-Yin Wang et. al

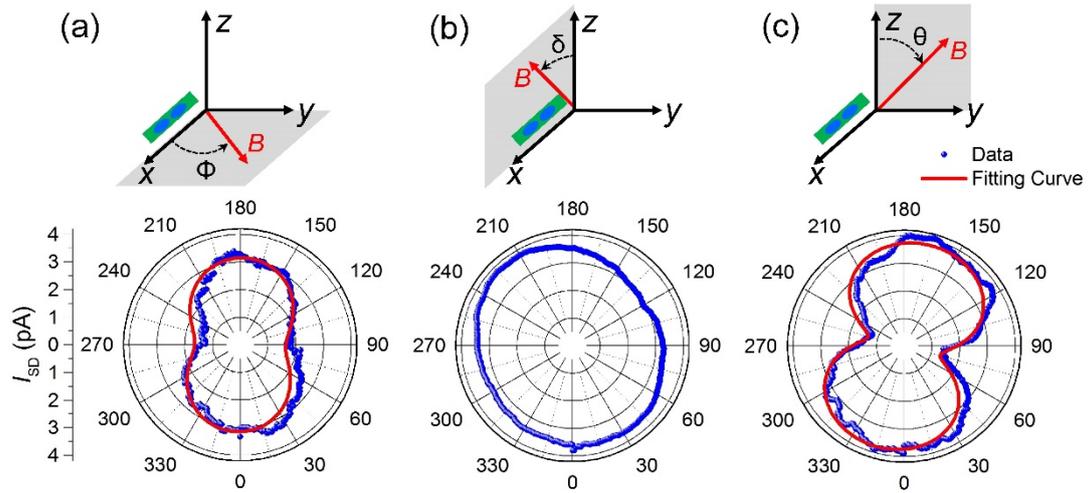

Fig. 4 by Ji-Yin Wang et. al

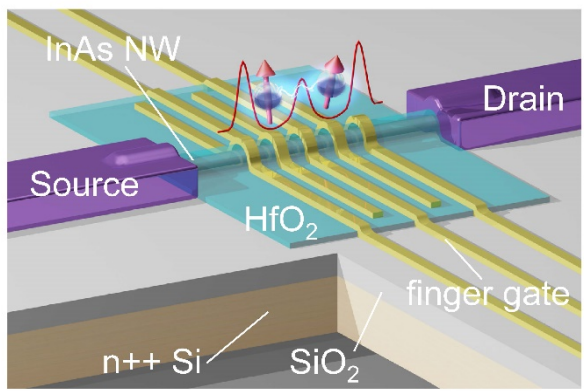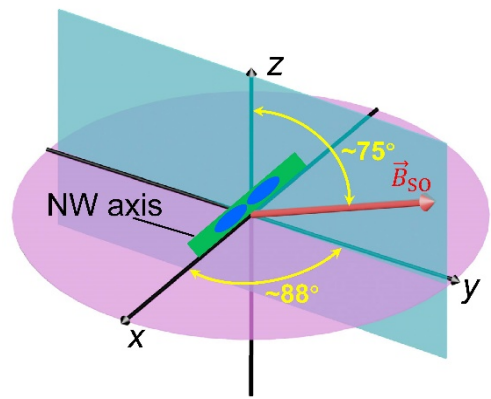

TOC figure by Ji-Yin Wang et. al

# Supporting Information for
# Anisotropic Pauli spin-blockade effect and spin-orbit interaction field in an InAs nanowire double quantum dot

Ji-Yin Wang,[†] Guang-Yao Huang,[†] Shaoyun Huang,*,[†] Jianhong Xue,[†] Dong Pan,[‡]
Jianhua Zhao,*,[‡] and H. Q. Xu*,[†,§]

[†] Beijing Key Laboratory of Quantum Devices, Key Laboratory for the Physics and Chemistry of Nanodevices, and Department of Electronics, Peking University, Beijing 100871, China

[‡] State Key Laboratory of Superlattices and Microstructures, Institute of Semiconductors, Chinese Academy of Sciences, Beijing 100083, China

[§] Division of Solid State Physics, Lund University, Box 118, S-22100 Lund, Sweden

*Emails: hqxu@pku.edu.cn; syhuang@pku.edu.cn; jhzhao@semi.ac.cn

## Abstract

In this Supporting Information, we provide more details and further discussion about the measurement results presented in the main article. We first present the charge stability diagram of the DQD measured in a large region of gate voltages at a large source-drain bias voltage of 30 mV. The measurements allow us to determine the few-electron DQD charge states including the empty electron (0,0) state. We then present the extraction of charging energies and orbital quantization energies of the two quantum dots (QDs) from the measured stability diagram of the DQD. Next, the procedure of the numerical fits to the measured leakage currents shown in Figures 3d, 3e and 3f of the main article are described in detail and the extracted results for the leakage current due to hyperfine interaction and spin-orbit interaction are presented and discussed. Finally, we present the magnetic-field-direction dependent measurements of the leakage current at a large detuning energy of $\varepsilon = 235$ μeV and discuss about that the presence of an off-substrate-plane angle in the direction of the spin-orbit field is a general feature in our nanowire device.

## I. Charge stability diagram of the DQD in a large region of gate voltages at a large source-drain bias voltage of 30 mV

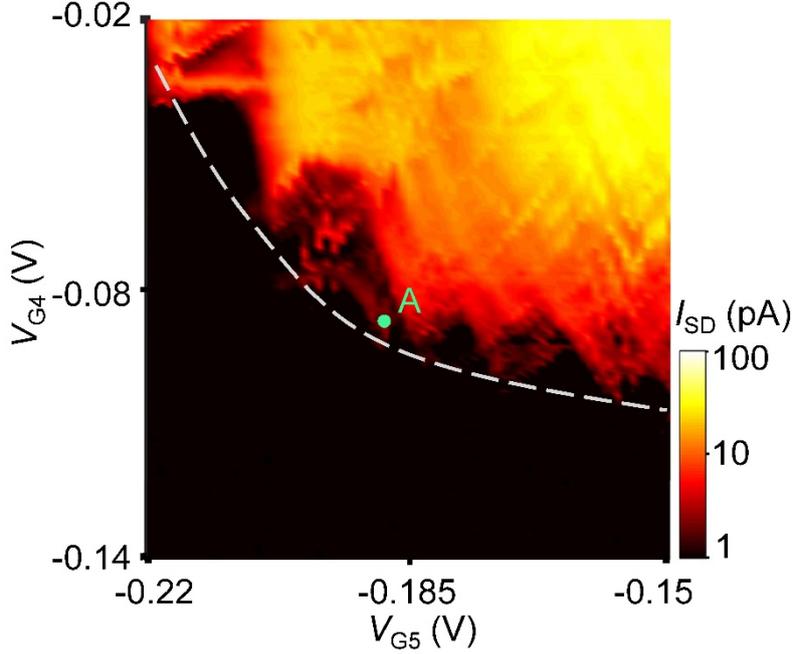

**Figure S1.** Source-drain current $I_{SD}$ of the DQD device studied in the main article measured as a function of $V_{G4}$ and $V_{G5}$ at a large source-drain bias voltage of $V_{SD}$ = 30 mV. The gray dashed curve separates the upper-corner region in which finite current $I_{SD}$ is observable from the region where no current is detectable even at this large source-drain bias voltage. The green point A marks the position where the DQD is at the start of its empty electron (0,0) state.

Figure S1 shows source-drain current $I_{SD}$ through the DQD as a function of $V_{G4}$ and $V_{G5}$ in a large gate voltage region at a large source-drain bias voltage of $V_{SD}$ = 30 mV. Finite current $I_{SD}$ is observed at the upper-right corner of the figure, i.e., the region above the gray dashed line. The green point A marks the position where the DQD is at the start of its empty electron (0,0) state. Here, a bias voltage of 30 mV applied is much larger than the addition energies of the two QDs (see the following section for the addition energies) and the empty (0,0) state is verified by the fact that no lower charge states are observed in the two QDs at this large source-drain bias voltage.

## II. Charging energies and orbital quantization energies of the left and right QDs

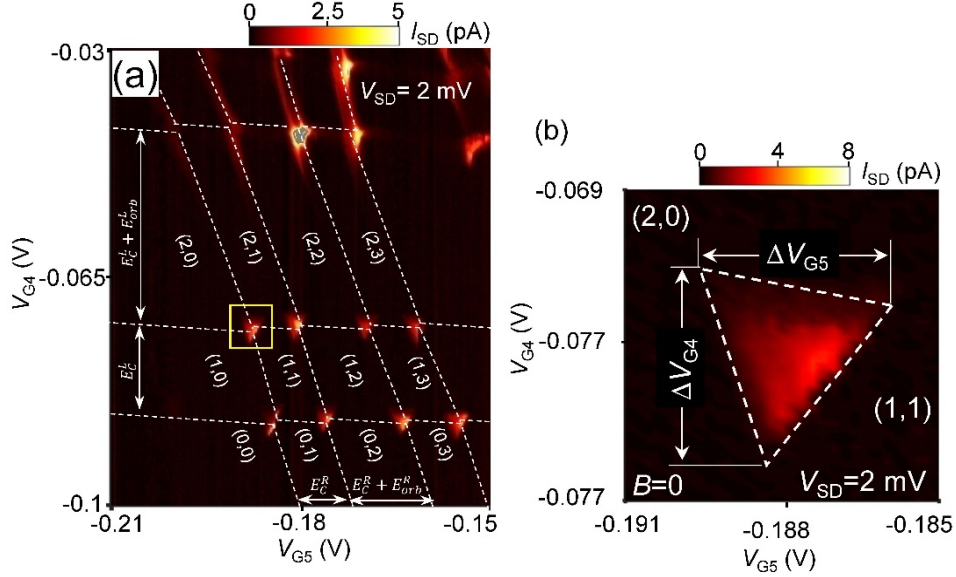

**Figure S2.** (a) The same charge stability diagram of the DQD as in Figure 1c in the main article. (b) Zoom-in plot of the measurements in the gate voltage region marked by the yellow rectangle in (a).

Here, we present the extraction of the charging energies and orbital quantization energies of the left and right QDs. Figure S2a show the same charge stability diagram of the DQD as in Figure 1c in the main article. The notation ($N_L$, $N_R$) indicates electron numbers on the left and right QDs. The white dashed lines denote the borders between the charge states. The energy required for adding an electron into the left (right) QD is proportional to the voltage spacing between two consecutive almost horizontal (vertical) dashed lines. Figure S2b is a zoom-in plot of the region marked by the yellow rectangle in Figure S2a. The sizes of the triangle, $\Delta V_{G4}$ and $\Delta V_{G5}$, indicated in Figure S2b are related to source-drain bias voltage $V_{SD}$ by $V_{SD} = \alpha_4 \times \Delta V_{G4} = \alpha_5 \times \Delta V_{G5}$. Therefore, we can extract the arm factors $\alpha_4 = 0.41$ and $\alpha_5 = 0.52$. With these arm factors, the charging energy $E_C^{L(R)}$ and the first orbital quantization energy $E_{orb}^{L(R)}$ in the left (right) QD are determined to be ~5.7 (~4.4) meV and ~6.5 (~2.1) meV, respectively.

## III. Discussion on leakage currents and numerical fittings

In this section, we give the details of numerical fits made that lead to the fitting results shown in Figures 3 of the main article. We observe finite leakage current through the DQD in the spin blockade regime due to spin state mixing. Previous works have demonstrated that hyperfine interaction and spin-orbit interaction are the two dominant mechanisms for the spin mixing in InAs nanowire DQDs.[1-3] The leakage currents due to these two spin mixing mechanisms show characteristic magnetic field dependences. Other spin mixing mechanisms, such as spin-flip co-tunneling, could also lead to observable leakage currents. However, these leakage currents can be assumed magnetic field independent. With all these considered, the leakage current $I_{\text{Leak}}$ through the DQD in the spin blockade regime at a finite magnetic field $|\vec{B}|$ can be estimated from,[4,5]

$$I_{\text{Leak}} = I_{\text{HF}}^0 S\left(\frac{\sqrt{3}|\vec{B}|}{B_{\text{N}}}\right) + I_{\text{SO}}^0 \left(1 - \frac{8B_{\text{C}}^2}{9\left(|\vec{B}|^2 + B_{\text{C}}^2\right)}\right) + I_{\text{B}}. \qquad (S1)$$

Here, the equation Eq. (S1) is the same as in Eq. (1) in the main article. The first term on the right side represents the hyperfine-induced leakage current, where $S(x)$ is a special function defined in Ref. 4 and $B_{\text{N}}$ is the effective nuclear field. The second term on the right side depicts the leakage current arising from spin-orbit interaction, which has a Lorentzian shaped dip with a width $B_{\text{C}}$, where $B_{\text{C}}$ is not but related to the strength of spin-orbit interaction.[5] The last term $I_{\text{B}}$ on the right side is an overall background leakage current caused by all other spin mixing mechanisms.

In Figures 3d, 3e and 3f of the main article, numerical fittings made by using Eq. (1) are presented red lines. The extract parameters from the fittings are shown in Table S1. Here, we note that the parameters $B_{\text{N}}$ and $I_{\text{B}}$ given with grey background in Table S1 for the cases corresponding to Figures 3e and 3f are not free fitting parameters but are taken from the values extracted for Figure 3d. This is because we observe both a current peak due to hyperfine interaction and a current dip due to spin-orbit interaction in Figure 3d, while in Figures 3e and 3f only dip shapes of leakage current are observed. Thus, the leakage current arising from hyperfine interaction is much smaller than the leakage currents induced by spin-orbit interaction in Figures 3e and 3f. Consequently, it is not accurate to extract the effective nuclear field $B_{\text{N}}$ from Figures 3e and 3f. But, instead, due to the fact that the effect of nuclear field on the leakage current is isotropic, the same $B_{\text{N}}$ value extracted from Figure 3d is used in the numerical fittings for Figures 3e and 3f. Similarly, since background leakage current should also be very small and magnetic-field-direction independent, the same value of $I_{\text{B}}$ extracted from Figure 3d is used in the numerical fittings for Figures 3e and 3f.

**Table S1.** Extracted parameters by numerical fittings of the measurements shown in Figures 3d, 3e and 3f in the main article to Eq. (S1). The parameters presented in the gray background are the values taken directly from that obtained for Figure 3d.

| Field orientation | $I_N^0$ (pA) | $B_N$ (mT) | $I_{SO}^0$ (pA) | $B_C$ (mT) | $I_B$ (pA) |
|---|---|---|---|---|---|
| Along the $y'$ axis (Figure 3d) | 2.9 | 5.6 | 2.2 | 16.5 | 0.3 |
| Along the $x'$ axis (Figure 3e) | 2.0 | 5.6 | 3.8 | 9.0 | 0.3 |
| Along the $z'$ axis (Figure 3f) | 2.5 | 5.6 | 4.5 | 8.5 | 0.3 |

### IV. Extraction of the direction of the spin-orbit field at detuning energy $\varepsilon = 235\ \mu eV$

In order to check the generality of the off-substrate-plane orientation of the spin-orbit field $\vec{B}_{SO}$, we have analyzed the leakage current through the DQD device measured at detuning energy $\varepsilon = 235\ \mu eV$ with the magnetic field rotated in three orthogonal planes as shown in Figure S3. We have observed similar results as shown in Figure 4 of the main article. In the *xy* and *yz* planes, the leakage current exhibits a strong magnetic-field-direction dependence, while the leakage current almost dose not change when the magnetic field is rotated in the *xz* plane. In the *xy* and *yz* planes, we have performed the same numerical fittings by using Eq. (2) in the main article and extracted the direction of the spin-orbit field. The spin-orbit field $\vec{B}_{SO}$ is found to pointing to a direction with an azimuth angle of ~92° from the *x* axis in the *xy* plane and a polar angle of ~80° from the *z* axis in the *yz* plane. Here again as we discussed in the main article, the ~2° off orientation of the spin-orbit field from the direction perpendicular to the nanowire in the *xy* plane is most likely due to uncertainties inevitable in the measurements and analysis. However, the off-substrate-plane angle of ~10° in the *yz* plane is found for the spin-orbit field, which is comparable to the result of ~15° obtained at $\varepsilon = 70\ \mu eV$ as presented in the main article. Thus, the presence of an off-substrate-plane angle in the direction of the spin-orbit field is a general feature in our nanowire device. A difference seen in the actual off-substrate-plane angle could arise from an increase in the Rashba spin-orbit field due to the changes made on the gate voltages giving this larger detuning energy.

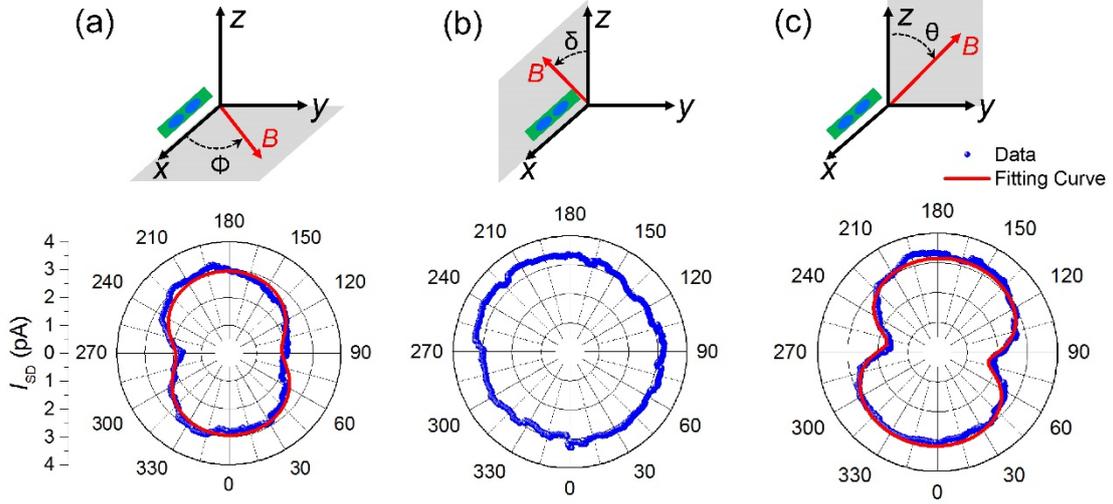

**Figure S3.** (a) Magnetic-field direction dependent measurements of source-drain current $I_{SD}$ with the applied magnetic field $\vec{B}$ rotated in the $xy$ plane as depicted in the top panel. $\phi$ is the angle between $\vec{B}$ and the $x$ axis in the $xy$ plane. (b) The same as in (a) but for $\vec{B}$ rotated in the $xz$ plane. $\delta$ is the angle between $\vec{B}$ and the $z$ axis in the $xz$ plane. (c) The same as in (a) but for $\vec{B}$ rotated in the $yz$ plane. $\theta$ is the angle between $\vec{B}$ and the $z$ axis in the $yz$ plane. In all the measurements, the nanowire is aligned with the $x$ axis, the strength of the magnetic field is fixed at 20 mT, and the detuning energy is fixed at 235 µeV. The red-solid lines in (a) and (c) depict the best fits to the theory Eq. (2) presented in the main article.